\documentclass[10pt]{article}
\usepackage{latexsym,graphicx}
\newcommand{\be}{\begin{equation}}
\newcommand{\ee}{\end{equation}}
\def\n{\noindent}
\catcode `\@=11
\catcode `\@=12
\begin{document}
\begin{center}
\large{\bf {LRS Bianchi Type-V Viscous Fluid Universe with a Time Dependent Cosmological Term $\Lambda$ }} \\
\vspace{10mm}
\normalsize{Anirudh Pradhan $^{a,}$\footnote{Corresponding Author}, J. P. Shahi $^{b}$ and 
Chandra Bhan Singh $^{c}$} \\
\vspace{2.5mm}
\normalsize{$^{a}$\it{Department of Mathematics, Hindu Post-graduate College, 
 Zamania-232 331, Ghazipur, India}} \\
{\it {E-Addresses: pradhan@iucaa.ernet.in, acpradhan@yahoo.com}}\\
\vspace{2.5mm}
\normalsize{$^{b,c}$\it{ Department of Mathematics, Harish Chandra Post-graduate College, 
Varanasi, India}}\\
\end{center}
\vspace{10mm}
\begin{abstract} 
An LRS Bianchi type-V cosmological models representing a viscous fluid distribution with a 
time dependent cosmological term $\Lambda$ is investigated. To get a determinate solution, 
the viscosity coefficient of bulk viscous fluid is assumed to be a power function of mass density. 
It turns out that the cosmological term $\Lambda(t)$ is a decreasing function of time, which is 
consistent with recent observations of type Ia supernovae. Various physical and kinematic features 
of these models have also been explored.
\end{abstract}
\smallskip
\n PACS number: 98.80.Es, 98.80.-k\\
\n Key words: cosmology, variable cosmological constant, viscous universe\\
%\newpage
%%%%%%%%%%%%%%%%%%%%%%%%%%%%%%%%%%%%%%%%%%%%%%%%%%%%%%%%%%%%%%%%%%%%%%%%%%%%%%%%%%%
%%%%%%%%%%%%%%%%%%%%%%%%%%%%%%%   SECTION 1  %%%%%%%%%%%%%%%%%%%%%%%%%%%%%%%%%%%%%%
\section{Introduction}
\noindent
Cosmological models representing the early stages of the Universe have been studied by 
several authors. An LRS (Locally Rotationally Symmetric) Binachi type-V spatially homogeneous 
space-time creates more interest due to its richer structure both physically and geometrically 
than the standard perfect fluid FRW models. An LRS Bianchi type-V universe is a simple 
generalization of the Robertson-Walker metric with negative curvature. Most cosmological models 
assume that the matter in the universe can be described by 'dust' (a pressure-less distribution) 
or at best a perfect fluid. However, bulk viscosity is expected to play an important role at 
certain stages of expanding universe \cite{ref1}$-$\cite{ref3}. It has been shown that bulk
viscosity leads to inflationary like solution \cite{ref4} and acts like a negative
energy field in an expanding universe \cite{ref5}. Furthermore, there are several
processes which are expected to give rise to viscous effects. These are the decoupling
of neutrinos during the radiation era and the decoupling of radiation and matter 
during the recombination era. Bulk viscosity is associated with the Grand Unification 
Theories (GUT) phase transition and string creation. Thus, we should consider the presence of 
a material distribution other than a perfect fluid to have realistic cosmological models 
(see Gr\o n \cite{ref6} for a review on cosmological models with bulk viscosity). A number of 
authors have discussed cosmological solutions with bulk viscosity in various context 
\cite{ref7}$-$\cite{ref9}.  
\newline
\par
Models with a relic cosmological constant $\Lambda$ have received considerable 
attention recently among researchers for various reasons 
(see Refs.{\cite{ref10}}$-${\cite{ref14}} and references therein). Some of the 
recent discussions on the cosmological constant ``problem'' and consequence on cosmology 
with a time-varying cosmological constant by Ratra and Peebles \cite{ref15}, 
Dolgov \cite{ref16}$-$\cite{ref18} and Sahni and Starobinsky \cite{ref19} have
pointed out that in the absence of any interaction with matter or radiation, the 
cosmological constant remains a ``constant''. However, in the presence of
interactions with matter or radiation, a solution of Einstein equations and the 
assumed equation of covariant conservation of stress-energy with a time-varying 
$\Lambda$ can be found. For these solutions, conservation of energy requires 
decrease in the energy density of the vacuum component to be compensated by a 
corresponding increase in the energy density of matter or radiation. Earlier 
researchers on this topic, are contained in Zeldovich \cite{ref20}, 
Weinberg \cite{ref11} and Carroll, Press and Turner \cite{ref21}. Recent
observations by Perlmutter et al. \cite{ref22} and Riess et al. \cite{ref23}
strongly favour a significant and positive value of $\Lambda$. Their finding arise from 
the study of more than $50$ type Ia supernovae with redshifts in the range
$0.10 \leq z \leq 0.83$ and these suggest Friedmann models with negative pressure
matter such as a cosmological constant $(\Lambda)$, domain walls or cosmic strings (Vilenkin
\cite{ref24}, Garnavich et al. \cite{ref25}) Recently, Carmeli and Kuzmenko \cite{ref26}
have shown that the cosmological relativistic theory (Behar and Carmeli \cite{ref27})
predicts the value for cosmological constant $\Lambda = 1.934\times 10^{-35} s^{-2}$.
This value of ``$\Lambda$'' is in excellent agreement with the measurements recently obtained 
by the High-Z Supernova Team and Supernova Cosmological Project (Garnavich et al.
\cite{ref25}, Perlmutter et al. \cite{ref22}, Riess  et al. \cite{ref23}, Schmidt 
et al. \cite{ref28}). The main conclusion of these observations is that the expansion 
of the universe is accelerating. 
\newline
\par
Several ans$\ddot{a}$tz have been proposed in which the $\Lambda$ term decays 
with time (see Refs. Gasperini \cite{ref29,ref30}, Berman \cite{ref31}, 
Freese et al. \cite{ref14}, $\ddot{O}$zer and Taha \cite{ref14}, 
Peebles and Ratra \cite{ref32}, Chen and Hu \cite{ref33}, Abdussattar and Viswakarma \cite{ref34},
Gariel and Le Denmat \cite{ref35}, Pradhan et al. \cite{ref36}). Of the special interest 
is the ans$\ddot{a}$tz $\Lambda \propto S^{-2}$ (where $S$ is the scale factor of the
Robertson-Walker metric) by Chen and Wu \cite{ref33}, which has been 
considered/modified by several authors ( Abdel-Rahaman \cite{ref37}, 
Carvalho et al. \cite{ref14}, Waga \cite{ref38}, Silveira and Waga \cite{ref39},
Vishwakarma \cite{ref40}).
\newline
\par
Recently Bali and Yadav \cite{ref41} obtained an LRS Bianchi type-V viscous fluid cosmological 
models in general relativity. Motivated by the situations discussed above, in this paper, 
we focus upon the exact solutions of Einstein's field equations in presence of a bulk 
viscous fluid in an expanding universe. We do this by extending the work of Bali
and Yadav \cite{ref41} by including a time dependent cosmological term $\Lambda$ in the field 
equations. We have also assumed the coefficient of bulk viscosity to be a power function of 
mass density. This paper is organized as follows. The metric and the field equations are presented 
in section 2. In section 3 we deal with the  solution of the field equations in presence of  
viscous fluid. The sections 3.1 and 3.2  contain the two different cases and also contain some 
physical aspects of these models respectively. Section $4$ describe two models under suitable 
transformations. Finally in section $5$ concluding remarks have been given.

%%%%%%%%%%%%%%%%%%%%%%%%%%%%%%%%%%%%%%%%%%%%%%%%%%%%%%%%%%%%%%%%%%%%%%%%%%%%%%%%%%%%%%%%%%%%
%%%%%%%%%%%%%%%%%%%%%%%%%%%%%%%  SECTION 2  %%%%%%%%%%%%%%%%%%%%%%%%%%%%%%%%%%%%%%%%%%%%%%%%
\section{The Metric and Field Euations}
We consider LRS Bianchi type-V metric in the form
\begin{equation} 
\label{eq1}  
ds^{2} = - dt^{2} + A^{2} dx^{2} + B^{2} e^{2x} (dy^{2} + dz^{2}),
\end{equation}
where A and B are functions of $t$ alone. \\

The Einstein's field equations (in gravitational units $c = 1$, $G = 1$) read as
\begin{equation} 
\label{eq2}
R^{j}_{i} - \frac{1}{2}Rg^{j}_{i} + \Lambda g^{j}_{i} = -8\pi T^{j}_{i}, 
\end{equation}
where $R^{j}_{i}$ is the Ricci tensor; $ R = g^{ij}R_{ij}$ is the Ricci scalar; and $T^{j}_{i}$ 
is the stress energy-tensor in the presence of bulk stress given by 
\[
T^{j}_{i} = (\rho + p)v_{i}v^{j} + p g^{j}_{i} - (v^{j}_{i;}+ v^{j}_{;i} + v^{j}v^{\ell}v_{i;\ell} 
+v_{i}v^{\ell}v^{j}_{;\ell})\eta
\]
\begin{equation} 
\label{eq3}
 - \left(\xi - \frac{2}{3}\eta \right)v^{\ell}_{;\ell}(g^{j}_{i} + v_{i}v^{j}).
\end{equation}
Here $\rho$, $p$, $\eta$ and $\xi$ are the energy density, isotropic pressure, coefficients of 
shear viscosity and bulk viscous coefficient respectively and $v^{i}$ the flow vector satisfying the 
relations 
\begin{equation} 
\label{eq4}
g_{ij}v^{i}v^{j} = -1.
\end{equation}
The semicolon $(;)$ indicates covariant differentiation. We choose the coordinates to be comoving, so 
that  $v^{i} = \delta^{i}_{4}$. \\

The Einstein's field equations (\ref{eq2}) for the line element (\ref{eq1}) has been set up as
\begin{equation} 
\label{eq5}
\frac{2B_{44}}{B} + \frac{B^{2}_{4}}{B^{2}} - \frac{1}{A^{2}} = - 8\pi \Big[p - 2\eta \frac{A_{4}}
{A} - \left(\xi - \frac{2}{3}\eta \right)\theta\Big] - \Lambda,
\end{equation}
\begin{equation} 
\label{eq6}
\frac{A_{44}}{A} + \frac{B_{44}}{B} + \frac{A_{4}B_{4}}{AB}  -\frac{1}{A^{2}} = - 8\pi \Big[p - 
2\eta \frac{B_{4}}{B} - \left(\xi - \frac{2}{3}\eta \right)\theta\Big] - \Lambda,
\end{equation}
\begin{equation} 
\label{eq7}
\frac{2A_{4}B_{4}}{AB} + \frac{B^{2}_{4}}{B^{2}} - \frac{3}{A^{2}} = - 8\pi \rho - \Lambda,
\end{equation}
\begin{equation} 
\label{eq8}
\frac{A_{4}}{A} - \frac{B_{4}}{B} = 0.
\end{equation}
The suffix 4 after the symbols $A$, $B$ denotes ordinary differentiation with respect to $t$ and 
$$\theta = v^{\ell}_{;\ell}
$$ 
%%%%%%%%%%%%%%%%%%%%%%%%%%%%%%%%%%%%%%%%%%%%%%%%%%%%%%%%%%%%%%%%%%%%%%%%%
%%%%%%%%%%%%%%%%%%%%%%%%%%  SECTION 3  %%%%%%%%%%%%%%%%%%%%%%%%%%%%%%%%%%
\section{Solutions of the Field Eqations}
In this section, we have revisited the solutions obtained by Bali and Yadav \cite{ref41}.
Equations (\ref{eq5}) - (\ref{eq8}) are four independent equations in seven  unknowns 
$A$, $B$, $p$, $\rho$, $\xi$, $\eta$ and $\Lambda$. For complete determinacy of the system, 
we need three extra conditions. 

Eq. (\ref{eq8}), after integration, reduce to
\begin{equation} 
\label{eq9}
A = B^{k},
\end{equation}
where $k$ is an integrating constant. Equations (\ref{eq5}) and (\ref{eq6}) lead to
\begin{equation} 
\label{eq10}
\frac{B_{44}}{B} - \frac{A_{44}}{A} - \frac{B^{2}_{4}}{B^{2}} - \frac{A_{4}B_{4}}{AB} 
= -16 \pi \eta \left(\frac{B_{4}}{B} - \frac{A_{4}}{A}\right).
\end{equation}
Using Eq. (\ref{eq9}) in (\ref{eq10}), we obtain
\begin{equation} 
\label{eq11}
\frac{df}{dB} + \left(\frac{k + 1}{B}\right)f = - 16\pi\eta,
\end{equation}
where $B_{4} = f(B)$. Eq. (\ref{eq11}) leads to
\begin{equation} 
\label{eq12}
f = - \frac{16\pi \eta}{(k + 2)} B + \frac{L}{B^{k + 1}},
\end{equation}
where $L$ is an integrating constant. Eq. (\ref{eq12}) again leads to
\begin{equation} 
\label{eq13}
B = (k + 2)^{\frac{1}{k + 2}}\left(k_{1} - k_{2}e^{-16\pi \eta t}\right)^{\frac{1}{k + 2}},
\end{equation}
where
\begin{equation} 
\label{eq14}
k_{1} = \frac{L}{16\pi \eta},
\end{equation}
\begin{equation} 
\label{eq15}
k_{2} = \frac{N}{16\pi \eta},
\end{equation}
N being constant of integration. From Eqs. (\ref{eq9}) and (\ref{eq13}), we obtain
\begin{equation} 
\label{eq16}
A = (k + 2)^{\frac{k}{k + 2}}\left(k_{1} - k_{2}e^{-16\pi \eta t}\right)^{\frac{k}{k + 2}}.
\end{equation}
Hence the metric (\ref{eq1}) reduces to the form
\[
ds^{2} = - dt^{2} + (k + 2)^{\frac{2k}{k + 2}}\left(k_{1} - k_{2}e^{-16\pi \eta t}\right)^{\frac{2k}
{k + 2}}dx^{2}
\]
\begin{equation} 
\label{eq17}
+ e^{2x} (k + 2)^{\frac{2}{k + 2}}\left(k_{1} - k_{2}e^{-16\pi \eta t}\right)^{\frac{2}{k + 2}}
(dy^{2} + dz^{2}).
\end{equation}
The pressure and density of the model (\ref{eq17}) are obtained as
\[
8\pi p = \frac{(8\pi)(16\pi \eta)k_{2}e^{-16\pi \eta t}}{3(k + 2)^{2}(k_{1} - k_{2}
e^{-16 \pi \eta t})^{2}}\Big[k_{1}(k + 2)^{2}(4\eta + 3\xi) - \{k^{2}(4\eta + 3\xi)  
\]
\begin{equation} 
\label{eq18}
+ 4k(\eta + 3\xi) + 2(5 \eta + 6\xi)\}k_{2}e^{-16 \pi \eta t}\Big] + \frac{1}{\left[(k + 2)
(k_{1} - k_{2}e^{-16 \pi \eta t})\right]^{\frac{2k}{k + 2}}} - \Lambda,
\end{equation}
\[
8\pi \rho = - \frac{(2k + 1)}{(k + 2)^{2}}(16 \pi \eta)^{2}k_{2}^{2}\frac{e^{-32\pi \eta t}}
{(k_{1} - k_{2}e^{-16 \pi \eta t})^{2}}
\]
\begin{equation} 
\label{eq19}
+ \frac{3}{\left[(k + 2)(k_{1} - k_{2}e^{-16 \pi \eta t})\right]^{\frac{2k}{k + 2}}} + \Lambda.
\end{equation}
The expansion $\theta$ in the model (\ref{eq17}) is obtained as
\begin{equation} 
\label{eq20}
\theta = \frac{(16 \pi \eta)k_{2}e^{-16 \pi \eta t}}{(k_{1} - k_{2}e^{-16 \pi \eta t})}.
\end{equation}
For complete determinacy of the system we have to consider three extra conditions. Firstly we 
assume that the coefficient of shear viscosity is constant, i.e., $\eta = \eta_{0}$ (say). For 
the specification of $\Lambda(t)$, we secondly assume that the fluid obeys an equation of state 
of the form
\begin{equation} 
\label{eq21}
p = \gamma \rho,
\end{equation}
where $\gamma(0 \leq \gamma \leq 1)$ is a constant.\\

Thirdly bulk viscosity $(\xi)$ is assumed to be a simple power function of the energy density 
\cite{ref42}$-$\cite{ref45}. 
\begin{equation}
\label{eq22}
\xi(t) = \xi_{0} \rho^{n},
\end{equation}
where $\xi_{0}$ and $n$ are constants. For small density, $n$ may even be equal to unity as used 
in Murphy's work \cite{ref46} for simplicity. If $n = 1$, Eq. (\ref{eq22}) may correspond
to a radiative fluid \cite{ref47}. Near the big bang, $0 \leq n \leq \frac{1}{2}$ is a more 
appropriate assumption \cite{ref48} to obtain realistic models. \\

For simplicity and realistic models of physical importance, we consider the following two cases 
$(n = 0, 1)$:
%%%%%%%%%%%%%%%%%%%%%%%%%%%%%%%%%%%%%%%%%%%%%%%%%%%%%%%%%%%%%%%%%%%%%%%%%
%%%%%%%%%%%%%%%%%%%%%%%%%%  SUBSECTION 3.1 %%%%%%%%%%%%%%%%%%%%%%%%%%%%%%%%%%
\subsection{Model I: ~ ~ ~ Solution for $n = 0$}
When $n = 0$, Eq. (\ref{eq22}) reduces to $\xi = \xi_{0}$ = constant. Hence, in this case Eqs. 
(\ref{eq18}) and (\ref{eq19}), with the use of (\ref{eq21}), lead to
\[
8\pi (1 + \gamma)\rho = \frac{8\pi M}{3N^{2}}\Big[k_{1}(k + 2)^{2}(4\eta_{0} + 3\xi_{0}) - 
\{k^{2}(4\eta_{0} + 3\xi_{0})  
\]
\begin{equation} 
\label{eq23}
+ 4k(\eta_{0} + 3\xi_{0}) + 2(5 \eta_{0} + 6\xi_{0})\}k_{2}e^{-16 \pi \eta_{0} t}\Big]  - 
\frac{(2k + 1)M^{2}}{N^{2}} + \frac{4}{N^{\frac{2k}{k + 2}}}.
\end{equation}
Eliminating $\rho(t)$ between Eqs. (\ref{eq19}) and (\ref{eq23}), we obtain
\[
(1 + \gamma)\Lambda  = \frac{8\pi M}{3N^{2}}\Big[k_{1}(k + 2)^{2}(4\eta_{0} + 3\xi_{0}) - 
\{k^{2}(4\eta_{0} + 3\xi_{0})  
\]
\begin{equation} 
\label{eq24}
+ 4k(\eta_{0} + 3\xi_{0}) + 2(5 \eta_{0} + 6\xi_{0})\}k_{2}e^{-16 \pi \eta_{0} t}\Big] + 
(2k + 1)\gamma\frac{M^{2}}{N^{2}} + \frac{(1 - 3\gamma)}{N^{\frac{2k}{k + 1}}},
\end{equation}
where
$$ M = 16\pi k_{2}\eta_{0}e^{-16\pi \eta_{0}t}, $$
$$ N = (k + 2)(k_{1} - k_{2}e^{-16\pi \eta_{0} t}),$$
$$ P = 2k^{2} + 2k + 5, $$
\begin{equation} 
\label{eq25}
Q = k^{2} + 4k + 4.
\end{equation}
%%%%%%%%%%%%%%%%%%%%%%%%%%%%%%%%%%%%%%%%%%%%%%%%%%%%%%%%%%%%%%%%%%%%%%%%%
%%%%%%%%%%%%%%%%%%%%%%%%%%  SUBSECTION 3.2 %%%%%%%%%%%%%%%%%%%%%%%%%%%%%%%%%%
\subsection{Model II: ~ ~ ~ Solution for $n = 1$}
When $n = 1$, Eq. (\ref{eq22}) reduces to $\xi = \xi_{0}\rho$ . Hence, in this case Eqs. 
(\ref{eq18}) and (\ref{eq19}), with the use of (\ref{eq21}), leads to
\[
8\pi \rho = \frac{16\pi M\{2k_{1}(k + 2)^{2}\eta_{0} - Pk_{2}\eta_{0}e^{-16\pi \eta_{0} t}\}}
{3\left[(1 + \gamma)N^{2} - M\{k_{1}(k + 2)^{2}\xi_{0} - Qk_{2}\xi_{0}e^{-16\pi \eta_{0} t}\}\right]}
\]
\begin{equation} 
\label{eq26}
+ \frac{4N^{\frac{4}{k + 2}} - (2k + 1)M^{2}}{\left[(1 + \gamma)N^{2} - M\{k_{1}(k + 2)^{2}\xi_{0} 
- Qk_{2}\xi_{0}e^{-16\pi \eta_{0} t}\}\right]}.
\end{equation}
Eliminating $\rho(t)$ between Eqs. (\ref{eq19}) and (\ref{eq26}), we get
\[
\Lambda = 16\pi M[2k_{1}(k + 2)^{2}\eta_{0} - Pk_{2}\eta_{0}e^{-16\pi \eta_{0} t}] + 
\frac{\gamma(2k + 1)}{(1 + \gamma)}\frac{M^{2}}{N^{2}} + \frac{(1 - 3\gamma)}{(1 + \gamma)N^{\frac{2k}
{k + 2}}} +
\]
\begin{equation} 
\label{eq27}
\frac{M[k_{1}(k + 2)^{2}\xi_{0} - Qk_{2}\xi_{0}e^{-16\pi \eta_{0} t}] \{4N^{\frac{4}{k + 2}} - 
(2k + 1)M^{2}\}}
{(1 + \gamma)N^{2}\left[(1 + \gamma)N^{2} - M\{k_{1}(k + 2)^{2}\xi_{0} - Qk_{2}\xi_{0}
e^{-16\pi \eta_{0} t}\}\right]}
\end{equation}
From Eqs. (\ref{eq23}) and (\ref{eq26}), we note that $\rho(t)$ is a decreasing function of time and 
$\rho > 0$ for all time in both models. The behaviour of the universe in these models will be 
determined by the cosmological term $\Lambda$; this term has the same effect as a uniform mass 
density $\rho_{eff} = - \Lambda/4\pi G$, which is constant in space and time. A positive value of 
$\Lambda$ corresponds to a negative effective mass density (repulsion). Hence, we expect that in the 
universe with a positive value of $\Lambda$, the expansion will tend to accelerate; whereas in the 
universe with negative value of $\Lambda$, the expansion will slow down, stop and reverse. From Eqs. 
(\ref{eq24}) and (\ref{eq27}), we observe that the cosmological term $\Lambda$ in both models is a 
decreasing function of time and it approaches a small positive value as time increase more and more. 
This is a good agreement with recent observations of supernovae Ia (Garnavich et al. \cite{ref25}, 
Perlmutter et al. \cite{ref22}, Riess  et al. \cite{ref23}, Schmidt et al. \cite{ref28}).\\

The shear $\sigma$ in the model (\ref{eq17}) is given by
\begin{equation} 
\label{eq28}
\sigma = \frac{(k - 1)M}{\sqrt{3}N}.
\end{equation} 
The non-vanishing components of conformal curvature tensor are given by
\begin{equation} 
\label{eq29}
C_{2323} = - C_{1414} = \frac{(k - 1)M}{3N^{2}}[kM - 16\pi\eta_{0}k_{1}(k + 2)],
\end{equation}  
\begin{equation} 
\label{eq30}
C_{1313} = - C_{2424} = \frac{(k - 1)M}{3N^{2}}[16\pi\eta_{0}k_{1}(k + 2) - kM],
\end{equation}  
\begin{equation} 
\label{eq31}
C_{1212} = - C_{3434} = \frac{(k - 1)M}{3N^{2}}[16\pi\eta_{0}k_{1}(k + 2) - kM].
\end{equation}  
Equations (\ref{eq20}) and (\ref{eq28}) lead to   
\begin{equation} 
\label{eq32}
\frac{\sigma}{\theta} = \frac{(k - 1)}{\sqrt{3}(k + 2)} = \mbox{constant}.
\end{equation}  
The model (\ref{eq17}) is expanding, non-rotating and shearing. Since $\frac{\sigma}{\theta} =$ 
conatant, hence the model does not approach isotropy. The space-time (\ref{eq17}) is Petrov 
type D in presence of viscosity.
%%%%%%%%%%%%%%%%%%%%%%%%%%%%%%%%%%%%%%%%%%%%%%%%%%%%%%%%%%%%%%%%%%%%%%%%%
%%%%%%%%%%%%%%%%%%%%%%%%%%  SECTION 4 %%%%%%%%%%%%%%%%%%%%%%%%%%%%%%%%%%
\section{Other Models}
After using the transformation
\begin{equation} 
\label{eq33}
k_{1} - k_{2}e^{-16\pi \eta t} = \sin{(16 \pi \eta \tau)}, ~ ~ ~ k + 2 = 1/16 \pi \eta,
\end{equation}
the metric (\ref{eq17}) reduces to
\[
ds^{2} = - \left[\frac{\cos{(16 \pi \eta \tau)}}{k_{1} - \sin{(16 \pi \eta \tau)}}\right]^{2}d\tau^{2}
+ \left[\frac{\sin{(16 \pi \eta \tau)}}{16 \pi \eta}\right]^{2(1 - 32\pi \eta)}dx^{2}
\]
\begin{equation} 
\label{eq34}
+ e^{2x} \left[\frac{\sin{(16 \pi \eta \tau)}}{16 \pi \eta}\right]^{(32\pi \eta)}(dy^{2} + dz^{2}).
\end{equation}
The pressure $(p)$, density $(\rho)$ and the expansion $(\theta)$ of the model (\ref{eq34}) are 
obtained as
\[
8\pi p = \frac{(16\pi \eta)^{2}\{k_{1} - \sin{(16\pi \eta \tau)}}{3\sin^{2}{(16\pi \eta \tau)}}
\Big[2k_{1} - 2(1 - 48\pi \eta + 1152 \pi^{2} \eta^{2})\{k_{1} - \sin{(16\pi \eta \tau)}\}\Big]
\]
\begin{equation} 
\label{eq35}
 + \frac{(16\pi \eta)(8\pi \xi)\{k_{1} - \sin{(16\pi \eta \tau)}\}}
{\sin{(16\pi \eta \tau)}} + \left[\frac{16 \pi \eta}{\sin{(16 \pi \eta \tau)}}\right]^
{2(1 - 32 \pi \eta)} - \Lambda,
\end{equation}
\begin{equation} 
\label{eq36}
8 \pi \rho =  \frac{2(24 \pi \eta - 1)(16\pi \eta)^{3}\{k_{1} - \sin{(16\pi \eta \tau)}\}^{2}}
{\sin^{2}{(16 \pi \eta \tau)}} + 3\left[\frac{16 \pi \eta}{\sin{(16 \pi \eta \tau)}}\right]^
{2(1 - 32 \pi \eta)} + \Lambda,
\end{equation}
\begin{equation} 
\label{eq37}
\theta = \frac{(16\pi \eta)\{k_{1} - \sin{(16 \pi \eta \tau)}\}}{\sin{(16\pi \eta \tau)}}.
\end{equation}
%%%%%%%%%%%%%%%%%%%%%%%%%%%%%%%%%%%%%%%%%%%%%%%%%%%%%%%%%%%%%%%%%%%%%%%%%
%%%%%%%%%%%%%%%%%%%%%%%%%%  SUBSECTION 4.1 %%%%%%%%%%%%%%%%%%%%%%%%%%%%%%%%%%
\subsection{Model I: ~ ~ ~ Solution for $n = 0$}
When $n = 0$, Eq. (\ref{eq22}) reduces to $\xi = \xi_{0}$ = constant. Hence, in this case Eqs. 
(\ref{eq35}) and (\ref{eq36}), with the use of (\ref{eq21}), lead to
\[
8\pi(1 + \gamma)\rho = \frac{2(16\pi\eta_{0})^{2} M_{1}}{3\sin^{2}(16\pi\eta_{0}\tau)}
\left[k_{1} - P_{1}M_{1}\right] + \frac{(16\pi \eta_{0})(8\pi \xi_{0})M_{1}}
{\sin(16\pi\eta_{0}\tau)} 
\]
\begin{equation} 
\label{eq38}
+ \, 4 N_{1} + \frac{2(24\pi \eta_{0} - 1)(16\pi \eta_{0})^{3}M^{2}_{1}}
{\sin^{2}(16\pi\eta_{0}\tau)}.
\end{equation}
Eliminating $\rho(t)$ between Eqs. (\ref{eq36}) and (\ref{eq38}), we obtain
\[
(1 + \gamma)\Lambda = \frac{2(16\pi\eta_{0})^{2} M_{1}}{3\sin^{2}(16\pi\eta_{0}\tau)}
\left[k_{1} - P_{1}M_{1}\right] + \frac{(16\pi \eta_{0})(8\pi \xi_{0})M_{1}}{\sin(16\pi\eta_{0}\tau)}
\]
\begin{equation} 
\label{eq39}
+ \, (1 - 3\gamma)N_{1} + \frac{2\gamma(24\pi \eta_{0} - 1)(16\pi \eta_{0})^{3}M^{2}_{1}}
{\sin^{2}(16\pi\eta_{0}\tau)}.
\end{equation}
%%%%%%%%%%%%%%%%%%%%%%%%%%%%%%%%%%%%%%%%%%%%%%%%%%%%%%%%%%%%%%%%%%%%%%%%%
%%%%%%%%%%%%%%%%%%%%%%%%%%  SUBSECTION 4.2 %%%%%%%%%%%%%%%%%%%%%%%%%%%%%%%%%%
\subsection{Model II: ~ ~ ~ Solution for $n = 1$}
When $n = 1$, Eq. (\ref{eq22}) reduces to $\xi = \xi_{0} \rho$ . Hence, in this case Eqs. 
(\ref{eq35}) and (\ref{eq36}), with the use of (\ref{eq21}), lead to
\[
8\pi \rho = \frac{2(16\pi\eta_{0})^{2} M_{1}[(k_{1} - P_{1}M_{1}) + 3(24\pi \eta_{0} - 1)
(16\pi \eta_{0})M_{1}]}{3\sin(16\pi\eta_{0}\tau)[(1 + \gamma)
\sin(16\pi\eta_{0}\tau) - 16\pi \eta_{0}\xi_{0}M_{1}]} 
\]
\begin{equation} 
\label{eq40} 
+ \, \frac{4N_{1}\sin(16\pi \eta_{0}\tau)}{[(1 + \gamma)\sin(16\pi\eta_{0}\tau) - 16\pi \eta_{0}
\xi_{0}M_{1}]}.
\end{equation}
Eliminating $\rho(t)$ between Eqs. (\ref{eq36}) and (\ref{eq40}), we obtain
\[
\Lambda = \frac{2(16\pi\eta_{0})^{2}M_{1}(k_{1} - P_{1}M_{1})}{3\sin(16\pi \eta_{0} \tau)
[(1 + \gamma)\sin(16\pi\eta_{0}\tau) - 16\pi \eta_{0}\xi_{0}M_{1}]}
\]
\[
+ \, N_{1}\frac{[3(16\pi \eta_{0} \xi_{0})M_{1} + (1 - 3\gamma)\sin(16\pi \eta_{0}\tau)]}
{[(1 + \gamma)\sin(16\pi\eta_{0}\tau) - 16\pi \eta_{0}\xi_{0}M_{1}]} \, + 
\]
\begin{equation} 
\label{eq41} 
\frac{2(24\pi\eta_{0} - 1)(16\pi\eta_{0})^{3}M^{2}_{1}[\gamma(1 + \gamma)\sin(16\pi \eta_{0}\tau) - 
(1 - \gamma)(16\pi \eta_{0}\xi_{0})M_{1}]}{(1 + \gamma)\sin^{2}(16\pi \eta_{0}\tau)[(1 + \gamma)
\sin(16\pi\eta_{0}\tau) - (16\pi \eta_{0}\xi_{0})M_{1}]},
\end{equation}
where
$$ M_{1} = k_{1} - \sin(16\pi\eta_{0} \tau), $$
$$ N_{1} = \left[\frac{16 \pi \eta_{0}}{\sin{(16 \pi \eta_{0} \tau)}}\right]^{2(1 - 32 \pi \eta)}, $$
\begin{equation} 
\label{eq42}
P_{1} = 1 - 48\pi \eta_{0} + 1152 \pi^{2}\eta^{2}_{0}.  
\end{equation}
The shear $(\sigma)$ in the model (\ref{eq34}) is obtained as
\begin{equation} 
\label{eq43}
\sigma = \frac{(1 - 48 \pi \eta_{0})(16\pi \eta_{0})[k_{1} - \sin(16\pi\eta_{0} \tau)}{\sqrt{3}
\sin(16\pi\eta_{0} \tau)}.
\end{equation}
The models descibed in cases $4.1$ and $4.2$ preserve the same properties as in the cases of 
$3.1$ and $3.2$. 
%%%%%%%%%%%%%%%%%%%%%%%%%%%%%%%%%%%%%%%%%%%%%%%%%%%%%%%%%%%%%%%%%%%%%%%%%%%%%%%%%
%%%%%%%%%%%%%%%%%%%  SECTION 4  %%%%%%%%%%%%%%%%%%%%%%%%%%%%%%%%%%%%%%%%%%
\section {Conclusions}
We have obtained a new class of LRS Bianchi type-V cosmological models of the universe 
in presence of a viscous fluid distribution with a time dependent cosmological term 
$\Lambda$. We have revisited the solutions obtained by Bali and Yadav \cite{ref41} and 
obtained new solutions which also generalize their work.
\newline
\par
The cosmological constant is a parameter describing the energy density of the vacuum
(empty space), and a potentially important contribution to the dynamical history
of the universe. The physical interpretation of the cosmological constant as 
vacuum energy is supported by the existence of the ``zero point'' energy predicted 
by quantum mechanics. In quantum mechanics, particle and antiparticle pairs are 
consistently being created out of the vacuum. Even though these particles exist 
for only a short amount of time before annihilating each other they do give the 
vacuum a non-zero potential energy. In general relativity, all forms of energy 
should gravitate, including the energy of vacuum, hence the cosmological constant.
A negative cosmological constant adds to the attractive gravity of matter, therefore
universes with a negative cosmological constant are invariably doomed to re-collapse 
\cite{ref49}. A positive cosmological constant resists the attractive gravity of matter
due to its negative pressure. For most universes, the positive cosmological constant
eventually dominates over the attraction of matter and drives the universe to expand 
exponentially \cite{ref50}.
\newline
\par
The cosmological constants in all models given in Sections $3.1$ and $3.2$ are decreasing
functions of time and they all approach a small and positive value at late times
which are supported by the results from recent type Ia supernova 
observations recently obtained by the High-z Supernova Team and Supernova Cosmological
Project (Garnavich {\it et al.} \cite{ref25}, Perlmutter {\it et al.} \cite{ref22},
Riess {\it et al.} \cite{ref23}, Schmidt {\it et al.} \cite{ref28}). Thus, with our approach,
we obtain a physically relevant decay law for the cosmological term unlike other
investigators where {\it adhoc} laws were used to arrive at a mathematical expressions 
for the decaying vacuum energy. Our derived models provide a good agreement with the
observational results. We have derived value for the cosmological constant $\Lambda$
and attempted to formulate a physical interpretation for it.\\
  
\section*{Acknowledgements} 
The authors wish to thank the Harish-Chandra Research Institute, Allahabad, India, for 
providing facility where part this work was done. We also thank to Professor Raj Bali 
for his fruitful suggestions and comments in the first draft of the paper.\\
%\newline
\newline

\end{document}